\documentclass[preprint,12pt]{elsarticle}

\usepackage{amsmath,amsfonts}
\usepackage{algorithm}
\usepackage{algpseudocode}
\usepackage{array}
\usepackage[caption=false,font=normalsize,labelfont=sf,textfont=sf]{subfig}
\usepackage{textcomp}
\usepackage{stfloats}
\usepackage{url}
\usepackage{verbatim}
\usepackage{graphicx}
\usepackage{balance}
\usepackage{hyperref}
\newtheorem{definition}{Definition}
\usepackage{caption}
\usepackage{subcaption}

\journal{Journal of Parallel and Distributed Computing}
\begin{document}
\begin{frontmatter}


\title{Workflow decomposition algorithm for scheduling with quantum annealer-based hybrid solver
}

\author[label1]{Marcin Kroczek}
\ead{marcinkroczek00@gmail.com}
\author[label1,label2]{Justyna Zawalska}
\ead{jzawalska@agh.edu.pl}
\author[label1,label2]{Katarzyna Rycerz}
\ead{kzajac@agh.edu.pl}
\address[label1]{AGH University of Krakow, Institute of Computer Science, al. Mickiewicza 30, 30-059 Krakow, Poland}
\address[label2]{Academic Computer Center Cyfronet AGH, Nawojki 11 Street, 30-950 Krakow, Poland}

\begin{abstract}
 We introduce the Series-Parallel Workflow Decomposition (SP\-WD) heuristic algorithm for the Workflow Scheduling Problem (WSP) decomposition. We demonstrate that the SPWD algorithm facilitates the scheduling of large WSP instances with the hybrid D-Wave Constrained Quadratic Model solver, enabling the scheduling of instances that would otherwise exceed its capacity limitations. We also describe the accompanying execution environment used to obtain the results of the experiments with real-life workflow instances available in the WfCommons standardization initiative repository.
 
\end{abstract}
\begin{graphicalabstract}
\includegraphics[width=\textwidth]{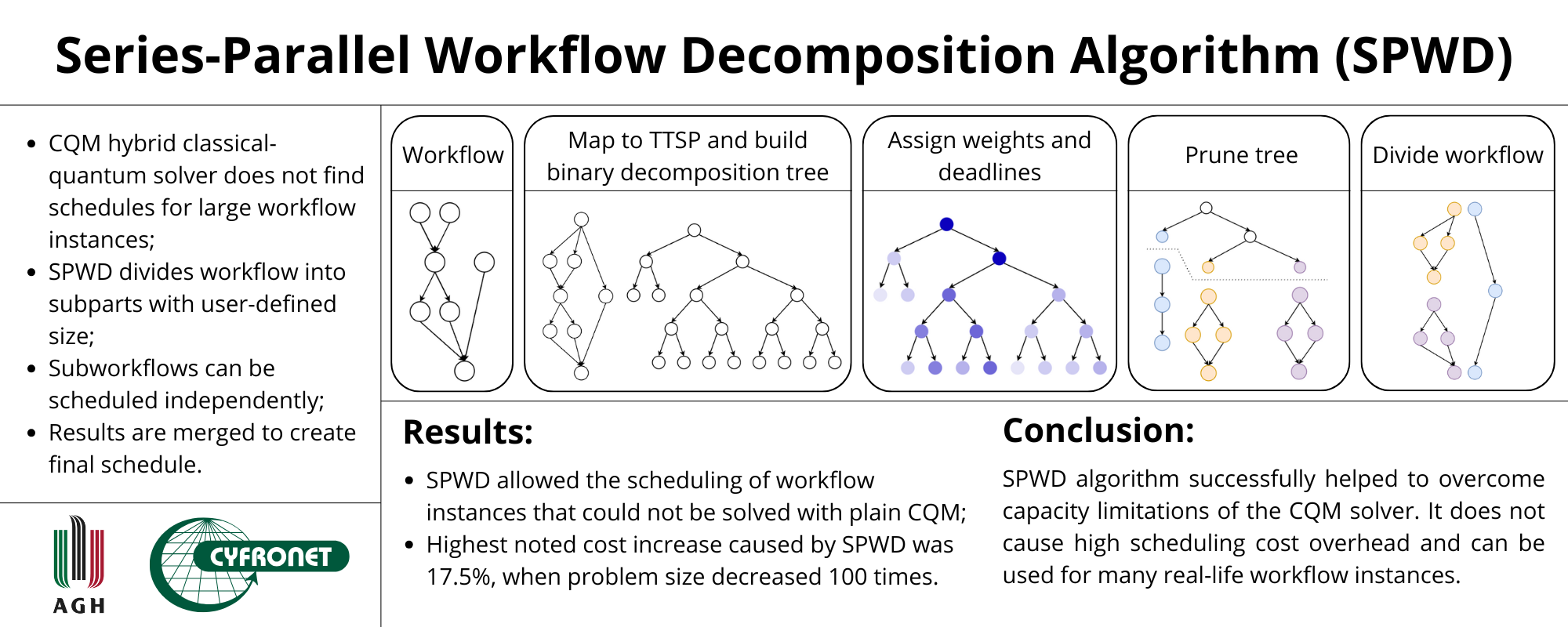}
\end{graphicalabstract}

\begin{highlights}
\item proposition of a new SPWD algorithm that enhances the capacity of the hybrid classical-quantum CQM solver when solving a workflow scheduling optimization problem;
\item experimental analysis of the algorithm  with real-life workflows data from the WfCommons repository;
\item a fully functional execution environment for performing aforementioned experiments.
\end{highlights}

\begin{keyword}
SPWD algorithm \sep Workflow scheduling \sep Quantum solvers \sep
CQM \sep WfCommons \sep QHyper
\end{keyword}
\end{frontmatter}


\section{Introduction}
A scientific workflow is an application model expressed as a Directed Acyclic Graph (DAG) of many tasks. Important problems in the field of astronomy~\cite{rynge_producing_2014}, bioinformatics~\cite{liu_pgen_2016}, high-energy physics~\cite{kilic_workflow_2024} or computational medicine~\cite{juve_characterizing_2013} can be expressed this way. The model is popular, so initiatives like the WfCommons framework~\cite{coleman_wfcommons_2022}  have been proposed to standardize the description of various real-life workflow applications. WfCommons is also a repository for  execution logs, which can be used as a reliable data source in research on the quality of workflow scheduling. 

Over the past few years, there have been several scientific publications focusing on the application of quantum computing to solve the problem of assigning workflow tasks to machines in a cloud environment, i.e., the Workflow Scheduling Problem. The foundations for research in this field were established in~\cite{krzhizhanovskaya_foundations_2020}, where schedules for small examples of WSP were successfully found using the D-Wave 2000Q pure quantum annealer. However, the proposed approach struggled to solve larger problem instances. Since the publication of this research, the capacity and architecture of quantum annealers have been improved. Furthermore, the supporting software has been expanded, in particular with the hybrid solver set provided by D-Wave~\cite{d-wave_d-wave_2024}. This led to further research~\cite{hurbol_mateusz_selected_2022} on solving WSPs using D-Wave's hybrid solver designed for problems expressed as a Constrained Quadratic Model (CQM)~\cite{d-wave_constraint_nodate}. The results were promising, but also indicated capacity limitations for larger problems. 

In this paper, we investigate the limitations of the hybrid CQM solver in more detail and propose a solution to overcome them and obtain meaningful results. To achieve this goal, we designed a new heuristic algorithm for the WSP decomposition called the Series-Parallel Workflow Decomposition (SPWD). The algorithm is based on the concept of series-parallel DAGs~\cite{valdes_recognition_1982} and is applied to produce reduced-size problems that are then solved independently by an external solver of choice, and finally the results are combined into the final schedule. Moreover, we integrated our algorithm with the QHyper library~\cite{lamza_qhyper_2024}. This allowed us to build a flexible environment for experiments that gets WfCommons workflow data and transforms it into a problem description that is accepted by both the hybrid classical-quantum CQM solver and the classical Gurobi optimizer~\cite{gurobi_optimization_gurobi_2024}, which we used as a reference method~\cite{lamza_qhyper_2024}.

Experimental results on the most popular families of real-life workflows available in the WfCommons repository show that the proposed SPWD algorithm combined with CQM finds schedules that were previously unattainable through the exclusive use of CQM. Additionally, we present the impact of both the number of binary variables and the amount of constraints on scheduling quality. Since the CQM solver is available for independent researchers as a black-box solution, such experiments are also important to investigate its implicit behavior.  Apart from the main research objective, the presented algorithm can also be used in strictly classical scenarios when the workflow scheduling problem is too large to be solved in a single attempt.

To summarize, our paper offers the following contributions:
\begin{itemize}
\item proposition of a new SPWD algorithm that enhances capacity of  hybrid classical-quantum CQM solver when solving workflow scheduling optimization problem;
\item experimental analysis of the algorithm  with real-life workflows data from WfCommons repository;
\item fully functional execution environment for performing aforementioned experiments.
\end{itemize}
The paper is organized as follows:  the necessary basic concepts can be found in Section~\ref{sec:preliminaries} and the overview of the related work is described in Section~\ref{sec:related}. Section~\ref{sec:algorithm} explains the SPWD algorithm in detail and Section~\ref{sec:complexity} presents its time complexity analysis. The execution environment is shown in Section~\ref{sec:qhyper}. The results are presented in Section~\ref{sec:results}. Section~\ref{sec:conclusions} contains the conclusion.

\section{Preliminaries}
\label{sec:preliminaries}
In this Section, we provide an overview of the key concepts necessary for understanding the solution presented in this paper. 

\begin{definition}[Workflow Scheduling Problem (WSP)]
\label{def:workflow}
    \leavevmode
The WSP~\cite{krzhizhanovskaya_foundations_2020} is defined with  the following input data:
\begin{itemize}
     \item DAG  $G$ with $t$ nodes (tasks) and $e$ edges that defines the structure of a workflow including the paths set $P$ of $p$ paths between the workflow's roots and leaves.
     \item Time matrix $\mathnormal{T}=[\tau_{i,j}]_{t\times m}$, where $t$ is the number of tasks and $m$ is the number of machines. This matrix contains information about the execution time of each task on each available machine.
    \item Machine cost vector $K=[k_{i}]_m$. Contains the price of using each machine for a unit of time. This vector, in connection with the time matrix, allows one to calculate the cost of running a task on a given machine, which can be represented as the matrix $\mathnormal{C}=[c_{i,j}]_{t \times m}$.
    \item Deadline $d$ is an integer that imposes the time limit for the execution of the workflow.
\end{itemize}
The goal is to assign tasks to machines in a way that minimizes the overall cost of running the workflow, given deadline restrictions.
\end{definition}

The common requirement when using pure quantum devices is to express optimization problems as Quadratic Unconstrained Binary Optimization (QUBO) functions~\cite{glover_quantum_2019}. To formulate QUBO for WSP, a cost function and its constraints have to be combined using the so-called penalty weights~\cite{krzhizhanovskaya_foundations_2020}. Setting that hyperparameter for an arbitrary WSP is a nontrivial problem. Therefore, for our experiments,  we chose a different solution, that is, the aforementioned CQM, a hybrid classical-quantum version of the D-Wave solver, since it allows us to provide constraints and the cost function separately, in the form of quadratic and linear functions of  binary, integer, or continuous variables~\cite{d-wave_constraint_nodate}. Unfortunately, due to the proprietary implementation,  the details of using a quantum processing unit by the D-Wave's CQM solver are hidden from the users.  The solution presented in this paper allows for investigation of the black-box CQM solver properties. 

\begin{definition}[WSP formulation for CQM]
\label{def:CQMform}
For each possible pair (task, machine), we set the binary variable $x_{i,j}=1$ if task $i$ is assigned to machine $j$, and $0$ otherwise.  
\begin{itemize}
\item The goal is to minimize the overall cost of a schedule:
\begin{equation} \label{eq:objective_function}
        \sum_{t} \sum_{m} c_{t,m} \cdot x_{t,m}
\end{equation}

\item In order to assure that each task is run on exactly one machine, the set of one-hot constraints is defined:
\begin{equation} \label{eq:machines_constraint}
        \forall_{t}[(\sum_{m} x_{t,m}) = 1]
\end{equation}

\item Deadline requirement. Each path between a root and a leaf in the workflow has to finish its execution before the deadline:
\begin{equation} \label{eq:deadline_constraint}
        \forall_{p \in P}[(\sum_{t \in p}\sum_{m} \tau_{t,m} \cdot x_{t,m}) \leq d]
\end{equation}
\end{itemize}
\end{definition}
For the purpose of this paper, this formulation was applied to both the CQM solver and the reference Gurobi solver.

\begin{figure}[htbp]
	\centering
	\subfloat[Series composition]{\includegraphics[width=0.46\textwidth]{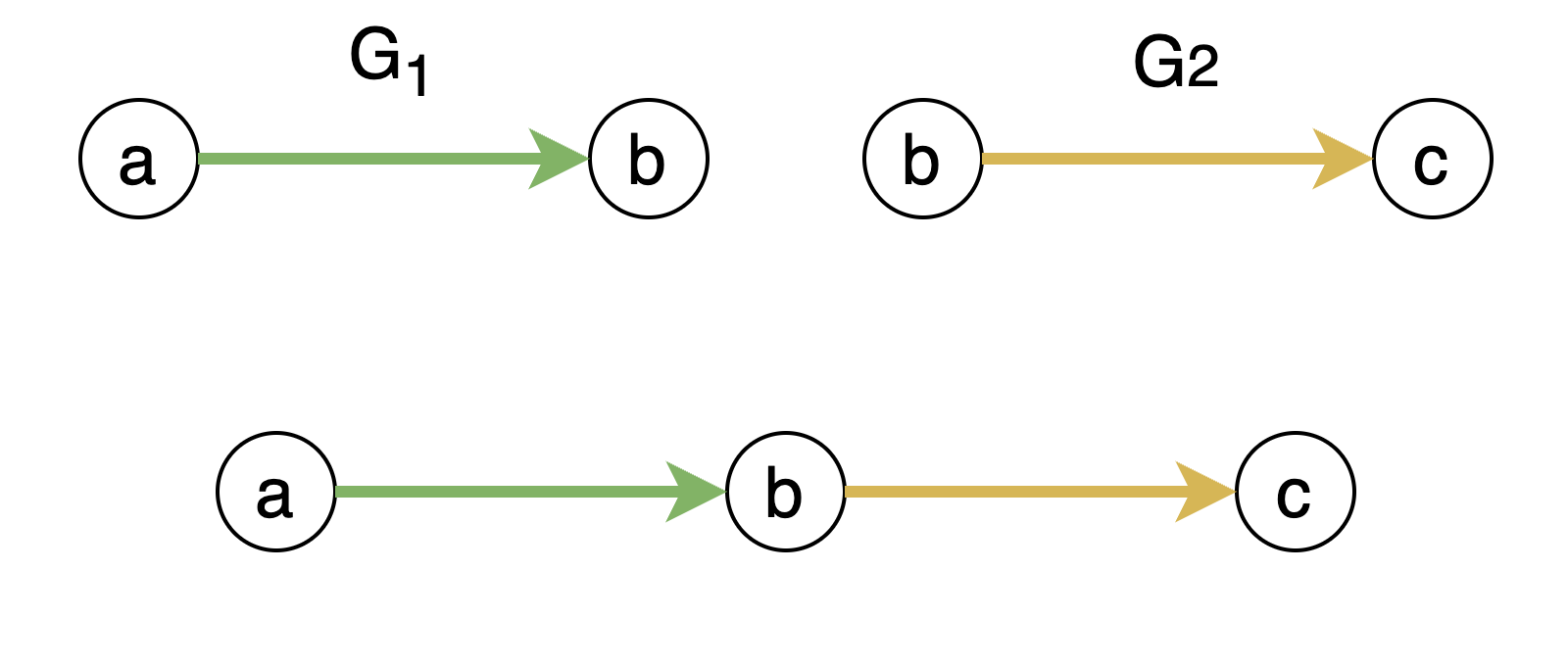}}
	\qquad
	\subfloat[Parallel composition]{\includegraphics[width=0.46\textwidth]{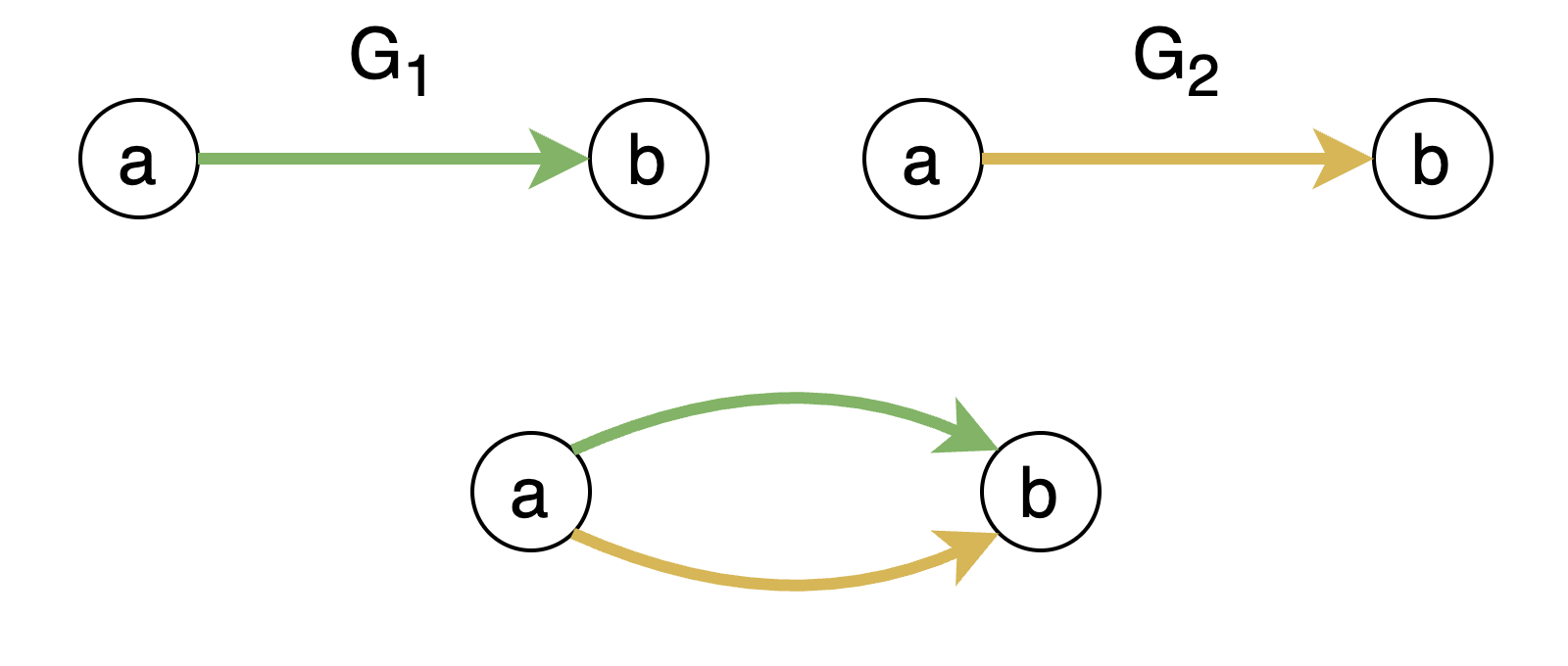}}
	\caption{Composition operations for TTSP multidigraphs.}
	\label{fig:ttsp_compositions} 
\end{figure}

\begin{figure}[h]
    \centering
    \includegraphics[width=0.7\textwidth]{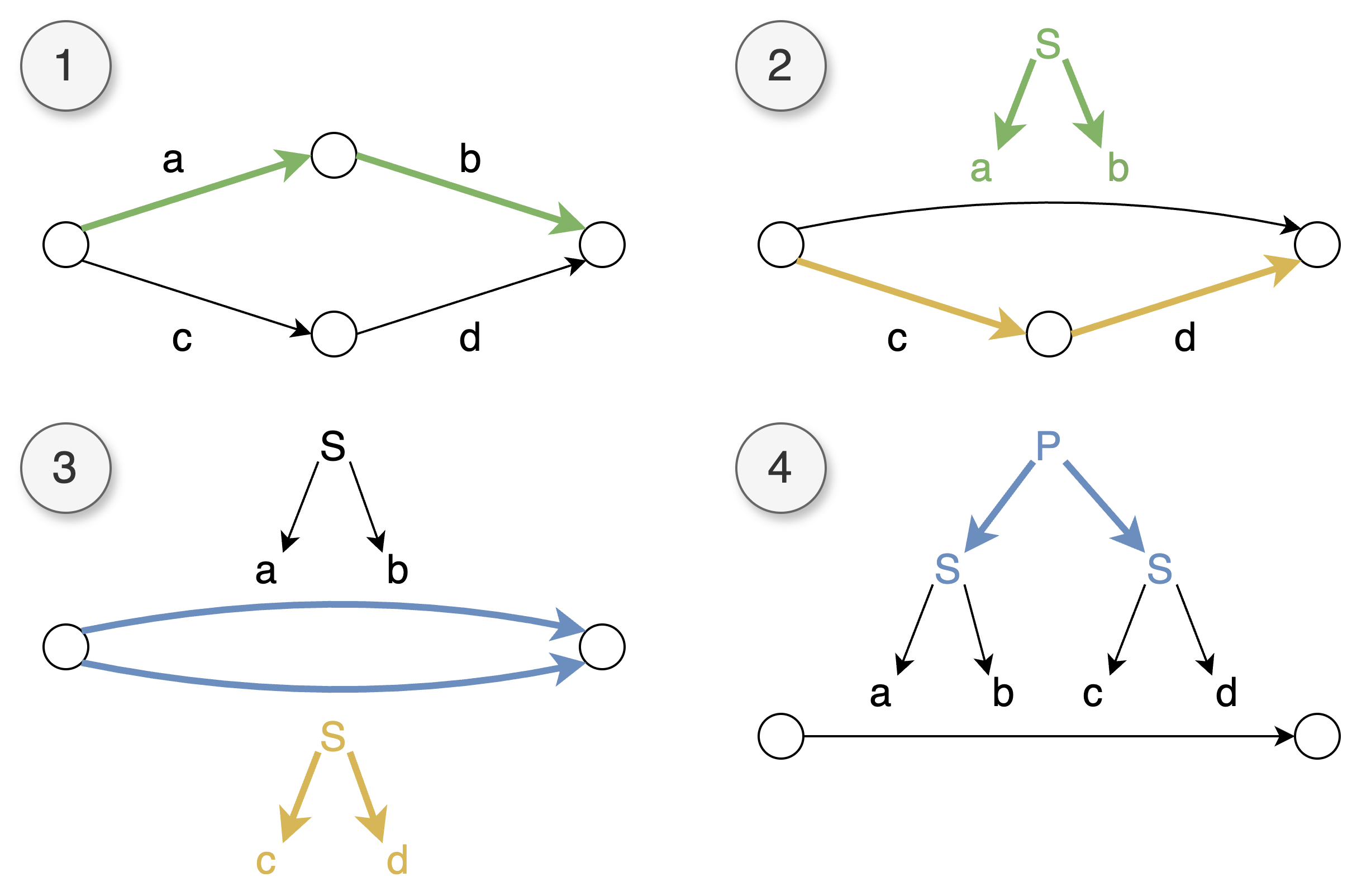}
    \caption{Process of building TTSP binary decomposition tree for graph with four vertices.}
    \label{fig:sp_tree}
\end{figure}

SPWD heavily relies on series-parallel DAGs, which constitute a group of workflows that are defined recursively using series and parallel compositions.
\begin{definition}[Two Terminal Series Parallel Multidigraphs (TTSP)\cite{valdes_recognition_1982}]
    \leavevmode
    \begin{enumerate}
        \item A directed graph consisting of two vertices joined by a single edge is TTSP. 
        \item If $G_1$ and $G_2$ are TTSP multidigraphs, so is the multidigraph obtained by either of the following operations: 
            \begin{enumerate}
                \item Two terminal series composition: identify the sink of $G_1$ with the source of $G_2$ (see Fig.~\ref{fig:ttsp_compositions}a).
                \item Two terminal parallel composition: identify the source of $G_1$ with the source of $G_2$ and the sink of $G_1$ with the sink of $G_2$ (see Fig.~\ref{fig:ttsp_compositions}b).
            \end{enumerate}
    \end{enumerate}
\end{definition}
Any TTSP multidigraph can be recognized in linear time, using the algorithm described in \cite{valdes_recognition_1982}. This method applies parallel and series compositions as long as it is possible and checks if the remaining graph consists of a single edge. The algorithm also allows building the so-called binary decomposition tree, that refers to the construction process of that graph (see Fig.~\ref{fig:sp_tree}). The leaves of such a tree represent the graph edges, and the nodes correspond to either parallel or series compositions. To construct a binary decomposition tree, at the beginning, each edge is associated with a trivial binary tree --- a single node. Merging trees is performed alongside series and parallel reductions as follows:
when edges $e_1$ and $e_2$ are reduced to a new edge $e_3$, create a new node labeled \emph{P} or \emph{S} respectively, set it as a parent for roots of trees of $e_1$ and $e_2$ and associate it with $e_3$.
The example is shown in Fig.~\ref{fig:sp_tree}. First step presents graph to be reduced. In the second step, series reduction (S) is performed with the $a$ and $b$ edges. The same procedure is repeated with the $c$ and $d$ edges in the third step. Finally, the two resulting edges are merged by parallel reduction (P).
It is worth mentioning that a single TTSP can be represented by many non-isomorphic decomposition trees, depending on the order of executed composition operations.

The SPWD algorithm presented in this work requires a DAG to TTSP mapping method.
In general, such methods include manual techniques applied to well-known graph structures, simple layering techniques, and more complex automatic algorithms~\cite{palma_mapping_2003}. In this work, we have used the solution presented in \cite{palma_mapping_2003}.

\section{Related Work}
\label{sec:related}
Workflow scheduling is an NP-hard problem~\cite{coffman_computer_1976}, with a wide range of approaches proposed in the literature over the years. An excellent classification of  supporting algorithms  was proposed in~\cite{masdari_towards_2016}. The first group contains heuristic solutions such as the popular Heterogeneous-Earliest-Finish-Time (HEFT)~\cite{topcuoglu_performance-effective_2002} or modern enhanced divide-and-conquer workflow scheduling (EDQWS) \cite{khojasteh_toussi_edqws_2022}.  The next group of metaheuristic algorithms uses nature-inspired methods such as Particle Swarm Optimization~\cite{pandey_particle_2010} or genetic algorithms~\cite{huang_workflow_2014}. There are also hybrid methods that combine both approaches~\cite{yassa_multiobjective_2013}. 

An important paradigm worth mentioning is the area-oriented DAG sche\-du\-ling~\cite{taufer_scheduling_2017}, which optimizes the average number of tasks eligible for execution at each step.
In particular, an interesting heuristic was presented in~\cite{cordasco_scheduling_2014}, where the authors proposed the area maximizing (A-M) algorithm for series-parallel DAGs. They constructed a binary decomposition tree for the workflow and explored it using the divide-and-conquer strategy. The goal was to generate the A-M schedule (topological sort of workflow nodes) for the whole graph, by recursively calculating it for subgraphs and merging results together. Conceptually, we applied a similar recursive tree exploration approach in the SPWD algorithm. However, both solutions have many differences, starting from the definition of a schedule, since we consider it as a mapping between tasks and machines. We also heavily rely on the workload associated with tasks, as opposed to~\cite{cordasco_scheduling_2014}, which was designed to be platform-agnostic. Finally, the scheduling goal is different since we minimize the execution cost under the deadline constraint.

Additionally, since using quantum devices for optimization problems has become a popular research topic, there are also approaches to formulate WSP as a QUBO problem, which can be solved with the D-Wave quantum annealer~\cite{krzhizhanovskaya_foundations_2020} or the IBM Qiskit simulator on gate-based quantum devices~\cite{plewa_variational_2021}. In ~\cite{hurbol_mateusz_selected_2022} the problem was later reformulated to CQM and solved using the D-Wave CQM hybrid solver. However, because of the architectural and functional limitations of quantum devices, the previously mentioned solutions cover only small problem samples. In this paper, we present a method for dealing with real-life instances from the established WfCommons repository~\cite{coleman_wfcommons_2022}.

\section{Series-Parallel Workflow Decomposition}
\label{sec:algorithm}
The proposed algorithm, based on TTSP multidigraphs,  is shown in Listing~\ref{alg:decomposition_algorithm}. Firstly, we build the TTSP graph from the input graph G (line 3) using the algorithm from~\cite{palma_mapping_2003} and the corresponding binary decomposition tree using the algorithm from~\cite{valdes_recognition_1982} (line 4). Next, we assign the weights (see Section~\ref{sec:weight_assignment}), modify selected series nodes (see Section~\ref{sec:modify_series_nodes}) and distribute the deadline (see Section~\ref{sec:deadline_distribution}) over the tree (lines 5-7). Then, we perform actual division according to the {\em max subgraph size} parameter (see Section~\ref{sec:devision}) (line 8), use an external solver of choice to find the schedule for all parts (lines 9-10) and merge the schedules (line 12). In the case of machine assignment collision between tasks shared by subworkflows, a faster machine returned by the scheduler is selected to ensure that the deadline is met in both subworkflows. 
\begin{algorithm}
\caption{The workflow decomposition algorithm}\label{alg:decomposition_algorithm}
\begin{algorithmic}[1]
\State Input $G$, $d$, $T$
\State Input {\em max subgraph size} parameter $(s)$ 
\State Build TTSP graph $G'$ from $G$  \Comment{algorithm~\cite{palma_mapping_2003}}
\State Build binary decomposition tree for $G'$  \Comment{algorithm~\cite{valdes_recognition_1982}}
\State Distribute weights over the tree \Comment{Section~\ref{sec:weight_assignment}}
\State Modify selected series nodes \Comment{Section~\ref{sec:modify_series_nodes}}
\State Distribute deadline over the tree \Comment{Section~\ref{sec:deadline_distribution}}
\State $Division \gets$ Divide $G'$, so that every subgraph has at most $s$ nodes \Comment{Section~\ref{sec:devision}}
\For{each $part$ in $Division$}
    \State Find schedule of $part$ with external solver of choice
\EndFor
\State Merge schedules into final schedule
\end{algorithmic}
\end{algorithm}

\subsection{Weight assignment}\label{sec:weight_assignment}
\begin{figure}[h]
    \centering
    \includegraphics[width=0.6\linewidth]{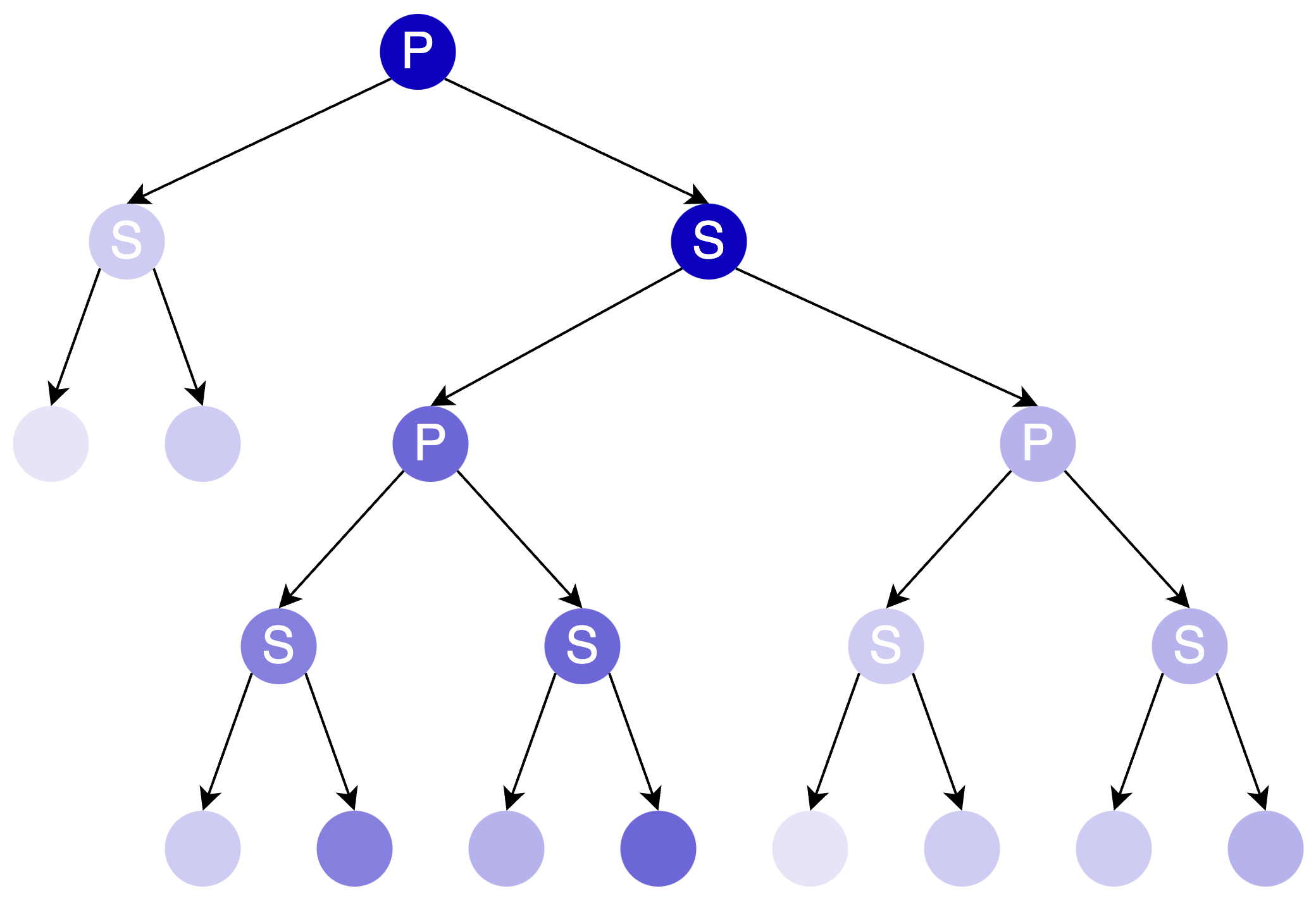}
    \caption{Binary decomposition tree with assigned weights. Darker colors represent nodes with bigger workloads.}
    \label{fig:weights_colors}
\end{figure}

The aim of this procedure is to assign weights to binary decomposition tree nodes, indicating a workload in a subtree rooted at that node. The tree root would accumulate workload from the whole graph, whereas the leaf would represent workload in a single edge. The weight assignment is based on the recursive tree traversal. Calculating the weight for a given node requires the weights from all of its children. The method starts from a tree root and recursively yields the children's weights. Weights are assigned according to the following policy:
\begin{enumerate}
    \item The weight $w(v)$ of a single node $v$ in the graph $G'$ is calculated as the mean execution time of this node across all possible machines:
    \begin{equation}
        w(v) = \frac{1}{m}\sum_{m}{\tau_{v,m}}
    \end{equation}
    \item The weight $w(e)$ of an edge $e = (v_1, v_2)$ is calculated as:
    \begin{equation}
        w(e) = w(v_1) + w(v_2)
    \end{equation}
    \item The weight of a node $n$ in the binary decomposition tree is calculated as: \begin{enumerate}
        \item If $n$ is a leaf associated with the edge $e$, then:
        \begin{equation}
            w(n) = w(e)
        \end{equation}
        \item If $n$ is a $series$ node, with children $c_1$ and $c_2$, connected by the graph vertex $v$, then:
        \begin{equation}
            w(n) = w(c_1) + w(c_2) - w(v)
        \end{equation}
        When subgraphs $c_1$ and $c_2$ are composed using $series$ operation, it means that they have common vertex $v$ that connects them. In this case, the weight of this vertex must be subtracted from the sum of the subgraph weights to ensure that it is not included twice.
        \item If $n$ is a $parallel$ node, with children $c_1$ and $c_2$, then:
        \begin{equation}
            w(n) = \max(w(c_1), w(c_2))
        \end{equation}
        In the case of $parallel$ composition of children $c_1$ and $c_2$, there exist two vertices $v_1$ and $v_2$, which are entry and exit tasks for both subgraphs. This composition allows to execute both branches in parallel (assuming access to an unlimited number of processors available on demand), so the final weight takes into account the weight indicating the heavier workload.
    \end{enumerate}
\end{enumerate}
An example binary decomposition tree with assigned weights is shown in Fig.~\ref{fig:weights_colors}. Darker colors represent nodes with bigger workloads.

\subsection{Deadline distribution} \label{sec:deadline_distribution}
Deadline should be proportional to the workload associated with a given decomposition tree node (see Section~\ref{sec:weight_assignment}). The deadline $d$ is assigned to the tree root and propagated down the tree. To ensure fair distribution, the following rules are applied:
\begin{enumerate}
    \item If node $n$ has associated deadline $d(n)$ and is a $series$ composition of children $c_1$ and $c_2$, distribute deadline over the children nodes, proportionally to their weights:
    \begin{equation} \label{eq:deadline_formula_series}
        d(c_i) = \frac{w(c_i)}{\sum_{i}w(c_i)}d(n) \quad i \in \{1, 2\}
    \end{equation}
    
    \item If node $n$ has associated deadline $d(n)$ and is $parallel$ composition of children $c_1$ and $c_2$, assign $d(n)$ to all children, because each branch must complete its execution under the same deadline.
    \begin{equation}
        d(c_i) = d(n) \quad i \in \{1, 2\}
    \end{equation}

    \item If the node is a leaf, there are no children to propagate the deadline.
\end{enumerate}

\subsection{Series nodes modification} \label{sec:modify_series_nodes}
\begin{figure}[htbp]
	\centering
	\subfloat[Connecting node with nonzero weight]{\includegraphics[width=0.42\linewidth]{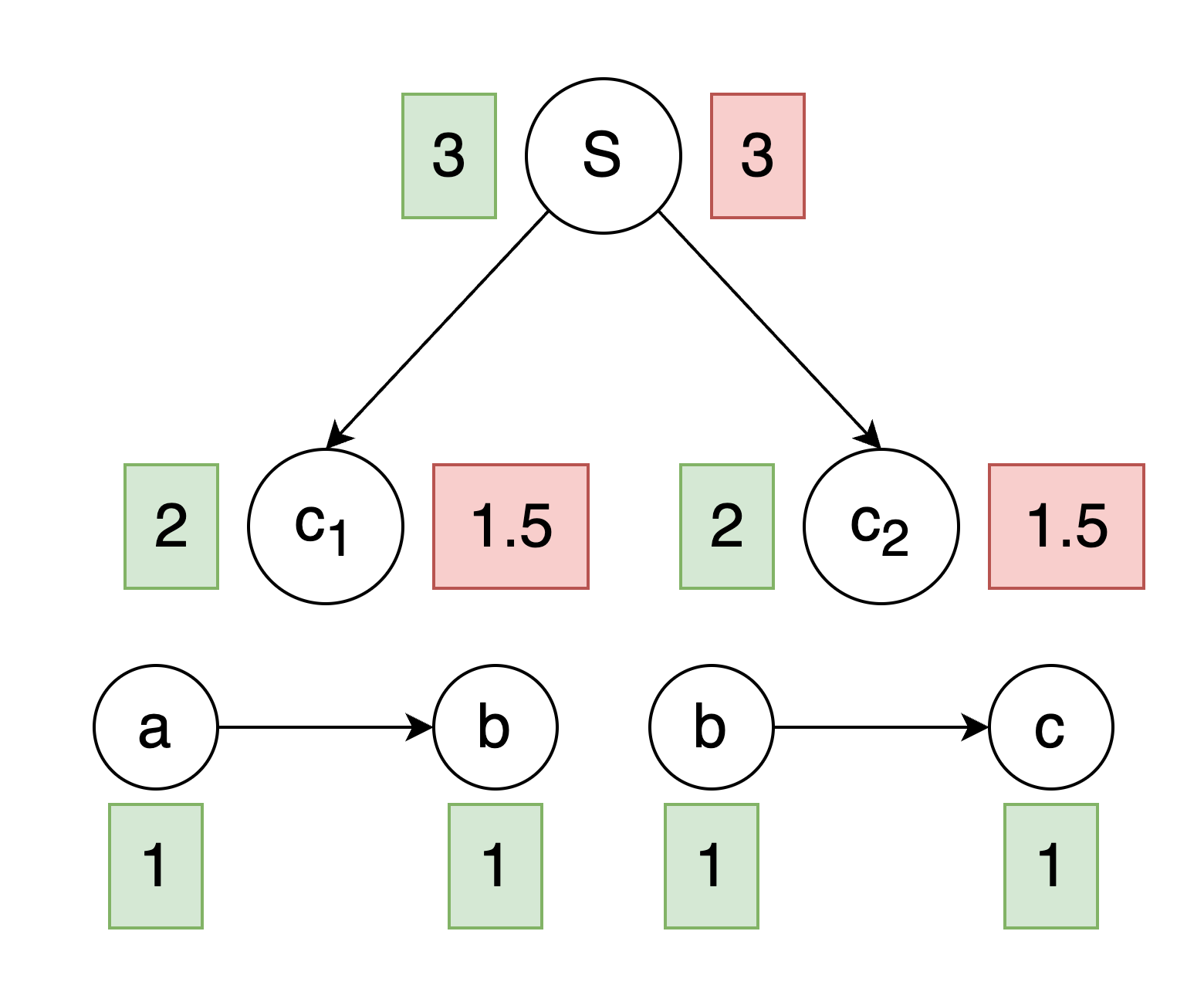}}
	\quad
	\subfloat[Connecting node with zero weight]{\includegraphics[width=0.42\linewidth]{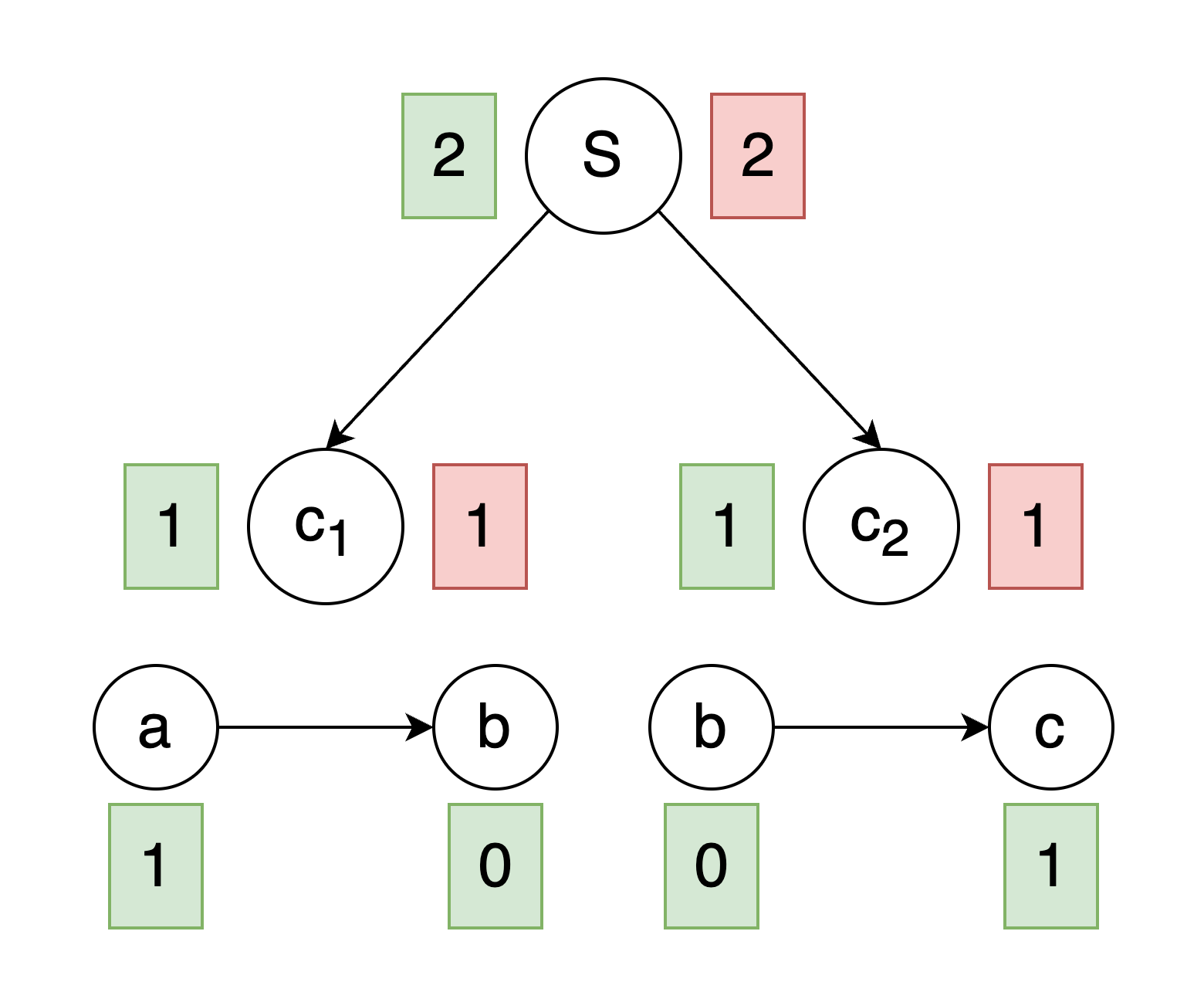}}
	\caption{Deadline distribution (red boxes) for series node, depending on the connecting node's weight (green boxes).}
	\label{fig:deadline_problem} 
\end{figure}

The deadline distribution for series nodes (described in Section~\ref{sec:deadline_distribution}) splits the deadline proportionally to the children's weights. Taking into account how the weights are calculated, the deadline is properly distributed only when the connecting vertex in the series node has zero weight. Fig.~\ref{fig:deadline_problem} shows how the deadline is distributed, depending on the weight of the connecting node. Weights are indicated by green color and deadlines by red. When the connecting node has a non-zero weight (Fig.~\ref{fig:deadline_problem}a) and we try to distribute the deadline with the value equal to the series node's weight (in this case $3$), the value $d = 1.5$ will be propagated to both children, according to Equation~\ref{eq:deadline_formula_series}. Note that we could not assign $d = 2$ to both of the children, because this could lead to a schedule that exceeds the deadline, e.g. $t_a = 1.9s$, $t_b=0.1s$ and $t_c=1.9s$ complies with the deadline for the children $d=2$, but exceeds $d=3$ for the series node. This issue does not occur when the connecting node has zero weight (Fig.~\ref{fig:deadline_problem}b), because both of the children get the deadline $d = 1$ as expected.

\begin{figure}[H]
    \centering
    \includegraphics[width=0.9\linewidth]{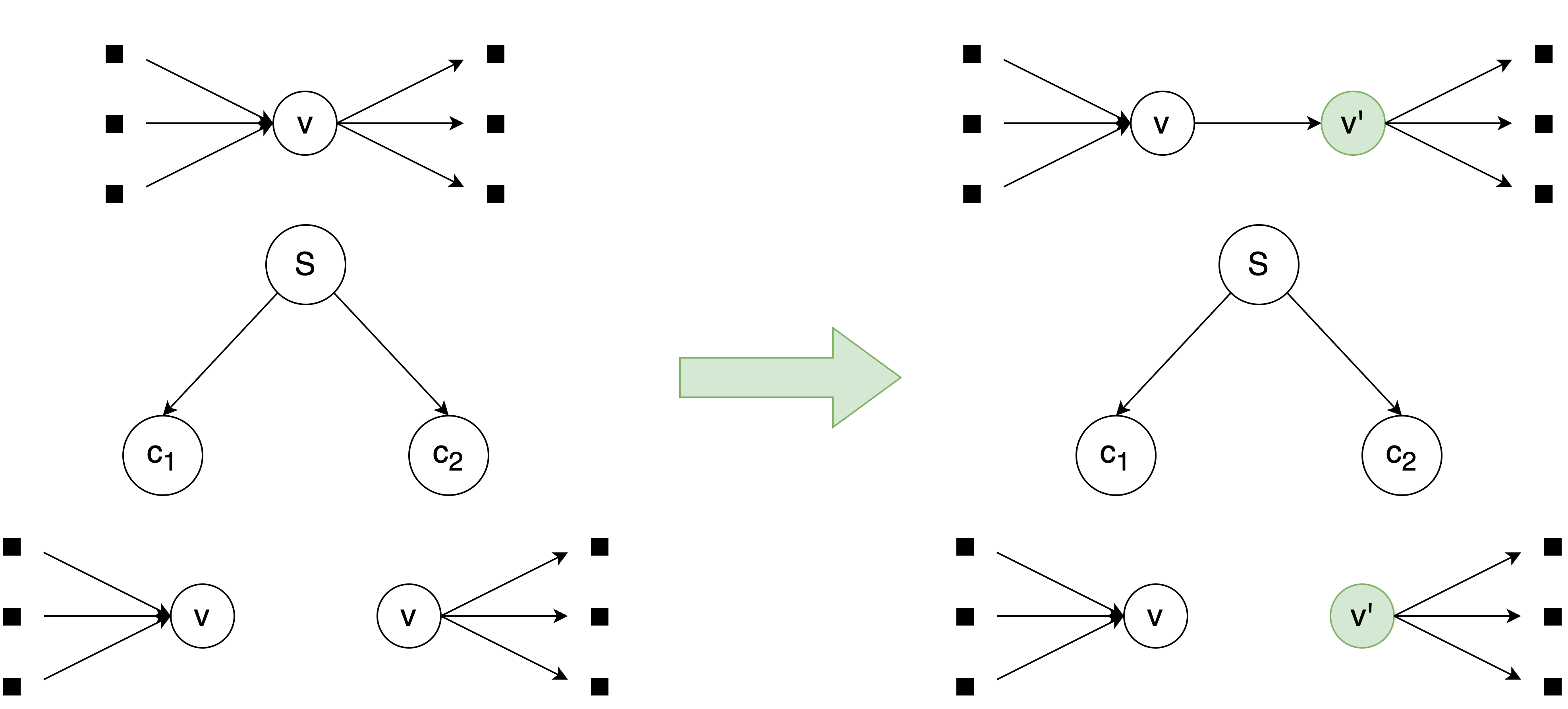}
    \caption{Series node modification using vertex substitution. New vertex $v'$ (green color) replaces $v$ in the right child's subgraph.}
    \label{fig:series_modification}
\end{figure}

To ensure that the connecting vertex for the series node will have zero weight, we introduce \emph{substitute vertices}. A substitute vertex for $v$ is denoted as $v'$, has zero weight, and is added to a workflow in a way that:
\begin{itemize}
    \item Outgoing edges $e = (v, u)$ are replaced with $e = (v', u)$;
    \item Entering edges $e = (u, v)$ remain unchanged;
    \item New edge $e = (v, v')$ is added.
\end{itemize} Substitute vertex must be created for each vertex having \textbf{nonzero weight}, being used as a connection in \textbf{series} node, whose children \textbf{will not be pruned} from the tree (see Section \ref{sec:devision}). The left child of the modified series node  remains unchanged (keeps the original vertex $v$), but in the right child $v'$  overrides $v$. This procedure is shown in Fig. \ref{fig:series_modification}.

\subsection{Workflow Division}
\label{sec:devision}
In this phase, an actual decomposition is executed based on the binary decomposition tree pruning. It is controlled by the {\em max subgraph size} parameter ($s$) and the goal is to split the graph into subgraphs with a number of vertices not exceeding $s$. As each node in the tree has a corresponding subgraph, information on the number of vertices can be easily extracted. The procedure is based on recursively checking the tree nodes and examining whether their subgraphs are small enough. If this is the case, the children of the node can be pruned. Otherwise, the procedure is continued for the node's children. When the procedure is finished, the tree leaves contain subgraphs whose size does not exceed $s$.

\section{Time Complexity Analysis}
\label{sec:complexity}
This section covers the time complexity analysis of the SPWD algorithm. It assumes that the graph is represented as an adjacency list and each tree node contains a list of represented graph vertices.

The mapping of the workflow to the TTSP multidigraph ($G'$) can be performed in $O(e + t \log t)$ using the algorithm from~\cite{palma_mapping_2003}. The number of nodes and edges may increase after mapping. According to~\cite{palma_mapping_2003}:
\begin{itemize}
    \item $t \leq t' \leq 2t$,
    \item $e \leq e' \leq 2(t'-2)$;
\end{itemize}
Building the decomposition tree according to~\cite{valdes_recognition_1982} runs in $O(e' + t')$ steps. The result is a full binary tree, thus the total number of its nodes is $2e'-1$, as $e'$ is equal to the number of leaves.

The weight assignment is a bottom-up procedure. Firstly, weights for tasks need to be calculated, which requires $t'm$ operations (to calculate the mean execution time for each task). Then, weights are calculated for each tree node ($2e'-1$). In summary, assigning weights has $O(e' + t'm)$ time complexity.

Modifying the series nodes can be implemented as a recursive procedure, starting from the tree root. It visits every node ($2e'-1$) exactly once and adds a substitute vertex if necessary, which takes constant time (modifying the pointers in the adjacency list and adding a new edge $v \rightarrow v'$). Along the recursion, nodes need to update their lists of vertices with substitute nodes, which is a linear operation. The total time complexity of this procedure is $O(e't')$.

Distributing the deadline is a recursive procedure that traverses the decomposition tree in a top-down manner. Each node is visited only once and executes a constant number of operations, which means that this phase runs in $O(e')$.

The workflow division starts with tree pruning, which in the worst case (no tree branch is removed) will walk through $2e'-1$ nodes. After pruning, the tree leaves contain subworkflows of the desired size. Extracting results can be performed in $O(e't')$ using a simple pointer manipulation, traversing the decomposition tree in a top-down manner. 

As mentioned above, the number of nodes and edges after TTSP mapping depends on the graph structure, thus the final time complexity is:
\begin{itemize}
    \item When $e'$ is closer to $e \rightarrow O(t(m + e + \log t))$;
    \item When $e'$ is closer to $2(t'-2)$ $\rightarrow O(tm + t^2)$.
\end{itemize}

\section{Execution Environment}
\label{sec:qhyper}

The SPWD algorithm is integrated with the QHyper library~\cite{lamza_qhyper_2024}, which provides the WSP implementation based on the common workflow format defined in the WfCommons framework. QHyper also offers a unified experimentation interface for tackling combinatorial optimization problems with pure quantum, hybrid classical-quantum, and classical solvers.

\begin{figure}[h!]
    \centering
    \includegraphics[width=0.7\linewidth]{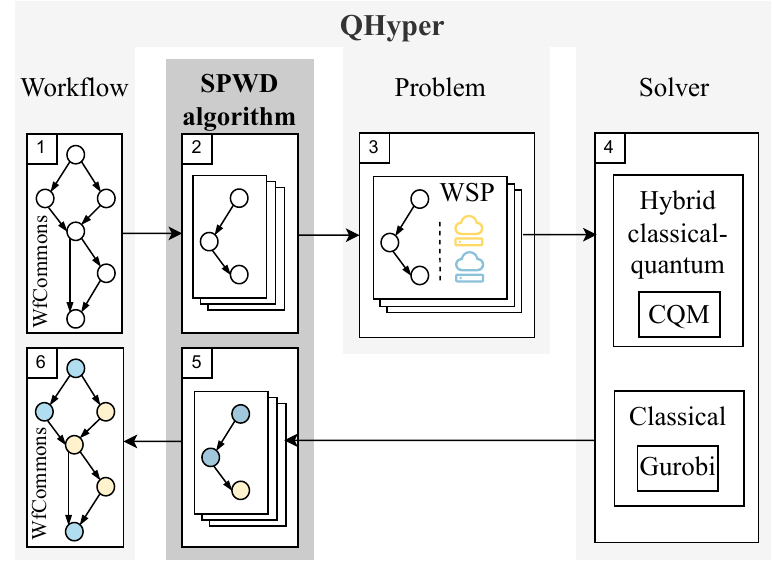}
    \caption{Execution steps of the SPWD algorithm interacting with the QHyper library.}
    \label{fig:execution_qrchitecture}
\end{figure}

The high-level interaction diagram between the SPWD implementation module and QHyper is presented in Fig.~\ref{fig:execution_qrchitecture}. It starts with the creation of a workflow instance with QHyper using a selected WfCommons real-life description in standardized JSON schema (step 1). Then, the SPWD algorithm module is used to transform this workflow into a collection of subworkflows as described in Section~\ref{sec:algorithm} (step 2). For each of the subworkflows, WSP description is created in QHyper through the assignment of a deadline and possible machines (step 3). Next, QHyper solves each of the subworkflows with a selected solver, either the hybrid classical-quantum CQM solver or the classical Gurobi optimizer (step 4). The results are then returned to the SPWD algorithm module (step 5), which merges them, resolves all the collisions, and ensures consistency throughout all the task-machine assignments (again, see Section~\ref{sec:algorithm}). Finally, the resulting schedule is returned, providing an optimized solution to the problem (step 6). 
The Python source code for SPWD is available in the GitHub repository\footnote{Repository containing the code of SPWD algorithm \url{https://github.com/mkroczek/WorkflowDecomposition}}.


\section{Results}
\label{sec:results}
The experiments described in this section were designed to explore various aspects of SPWD. All results from the experiments have been collected in a GitHub repository\footnote{Repository containing the results of experiments~\url{https://github.com/mkroczek/SPWD-experiments}}.

The aim of the first approach (Section~\ref{sec:mss_cost_experiment}) was to experimentally check the additional cost of the schedule, caused by the decomposition of a problem, in typical real-life examples from the WfCommons repository. To obtain the actual schedule, we used the Gurobi solver as it is a well-established classical software capable of solving all the workflow instances used. 

The second set of experiments (Section~\ref{sec:mss_problem_size}) tested the impact of decomposition on the number of constraints and variables in the problem. This part of the experiment did not require an actual solver as it checked only the decomposition algorithm itself.

The final set of experiments (Section~\ref{sec:mss_cqmcapasity}) examined if the capacity limitations of the classical-quantum CQM solver can be overcome by the SPWD algorithm, and at what schedule cost. The results were also compared with the results of the Gurobi solver.

\subsection {Methodology}
The real-life workflow instances used in the experiments include all workflow families from the WfInstances Pegasus  repository\footnote{WfInstances Pegasus repository~\url{https://github.com/wfcommons/WfInstances/tree/main/pegasus}}, i.e., 1000Genome~\cite{ferreira_da_silva_using_2019}, Epigenomics~\cite{juve_characterizing_2013}, SRA Search~\cite{nlm_sequence_nodate}, SoyKB~\cite{liu_pgen_2016}, Seismology~\cite{filgueira_asterism_2016}
and Montage~\cite{rynge_producing_2014}. 

For the experiments, we used a synthetic description of computing resources that contain $5$ machine types listed in Tab.~\ref{tab:cqm_machines}.
\begin{table}[H]
    \centering
    \caption{Machines used for scheduling in experiments. Speed is given in GHz, price is given per one second.}\label{tab:cqm_machines}
    \begin{tabular}{|c|c|c|c|}
        \hline
        name &speed [GHz] &core count &price per second \\
        \hline
        Machine1 &1.0 &5 &1.0 \\
        \hline
        Machine2 &1.25 &5 &1.5625 \\
        \hline
        Machine3 &1.5 &5 &2.25 \\
        \hline
        Machine4 &1.75 &5 &3.0625 \\
        \hline
        Machine5 &2.0 &5 &4.0 \\
        \hline
    \end{tabular}
\end{table}

The deadline for each workflow was set to a value that forced the scheduler to use mixed machine types. Its value differed between workflows and was set to the estimate of the time of the most consuming path in a workflow, i.e., the critical path value (CPV) of a given instance. More precisely, for each task in CPV we took its average execution time across all the machines.

The objective of scheduling is to minimize the cost under a certain deadline. Thus, the quality of the decomposed schedule can be measured as the increase in cost, relative to the cost of a schedule  without decomposition.

\subsection{Impact on the cost of the schedule}\label{sec:mss_cost_experiment}
The goal of this experiment was to measure how the {\em max subgraph size} parameter affects the
scheduling cost \eqref{eq:objective_function} for the selected workflow families (1000Genome, Epigenomics and SRA Search). Each family was represented by four workflow instances of various sizes. These workflows were divided into subgraphs according to the {\em max subgraph size}, expressed as a percentage of the total size of the workflow: $75\%, 50\%, 25\%, 15\%, 10\%, 5\%, 2\%, 1\%$. The problems were scheduled with the Gurobi solver. 

\begin{figure}
    \centering
    \includegraphics[width=0.8\linewidth]{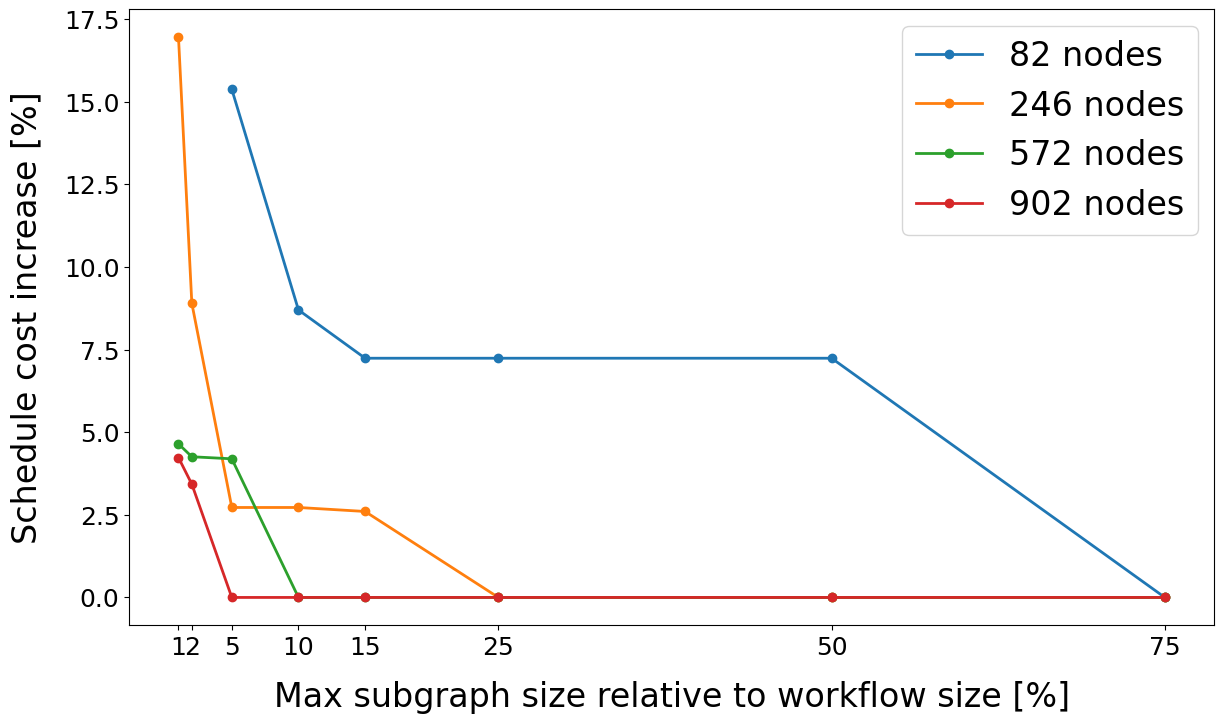}
    \caption{The {\em max subgraph size} influence on the scheduling cost for four workflow instances selected from the 1000Genome family.}
    \label{fig:mss_cost_increase_1000genome}
\end{figure}

The largest overhead was observed for the 1000Genome family, shown in Fig.~\ref{fig:mss_cost_increase_1000genome}. The cost of scheduling increases for small values of the maximum size of the subgraph and peaks up to $17.5\%$ for the workflow containing $82$ nodes, when the size decreased $100$ times. Larger instances tend to increase less than the smaller ones. Similar results, not shown in the figure, were observed for the Epigenomics family, with a maximum cost increase of $14\%$. The cost increase for SRA Search was static (maximum $2.5\%$) and did not change along with the parameter value.

\subsection{Impact on the problem size}
The purpose of this experiment was to determine how the SPWD algorithm affects the problem size, i.e. the number of necessary binary variables and constraints (see Def.~\ref{def:CQMform}) for known workflow families.
The question is not trivial, since
the relation between the {\em max subgraph size} parameter and the problem size depends on a specific workflow.
Moreover, the DAG to TTSP transformation may result in more paths and new vertices, and thus more constraints and variables.  

\subsubsection{Maximum subgraph size influence}\label{sec:mss_problem_size}
This section explores the relation between the {\em max subgraph size} and the problem size in divided subgraphs. 
The parameter value was sampled similarly to Experiment~\ref{sec:mss_cost_experiment}.
This experiment measured the relative change in the number of constraints and variables between the original graph and the largest divided subgraph (in terms of problem size). 
\begin{figure}[htbp]
	\centering
	\subfloat[1000Genome]{\includegraphics[width=\linewidth]{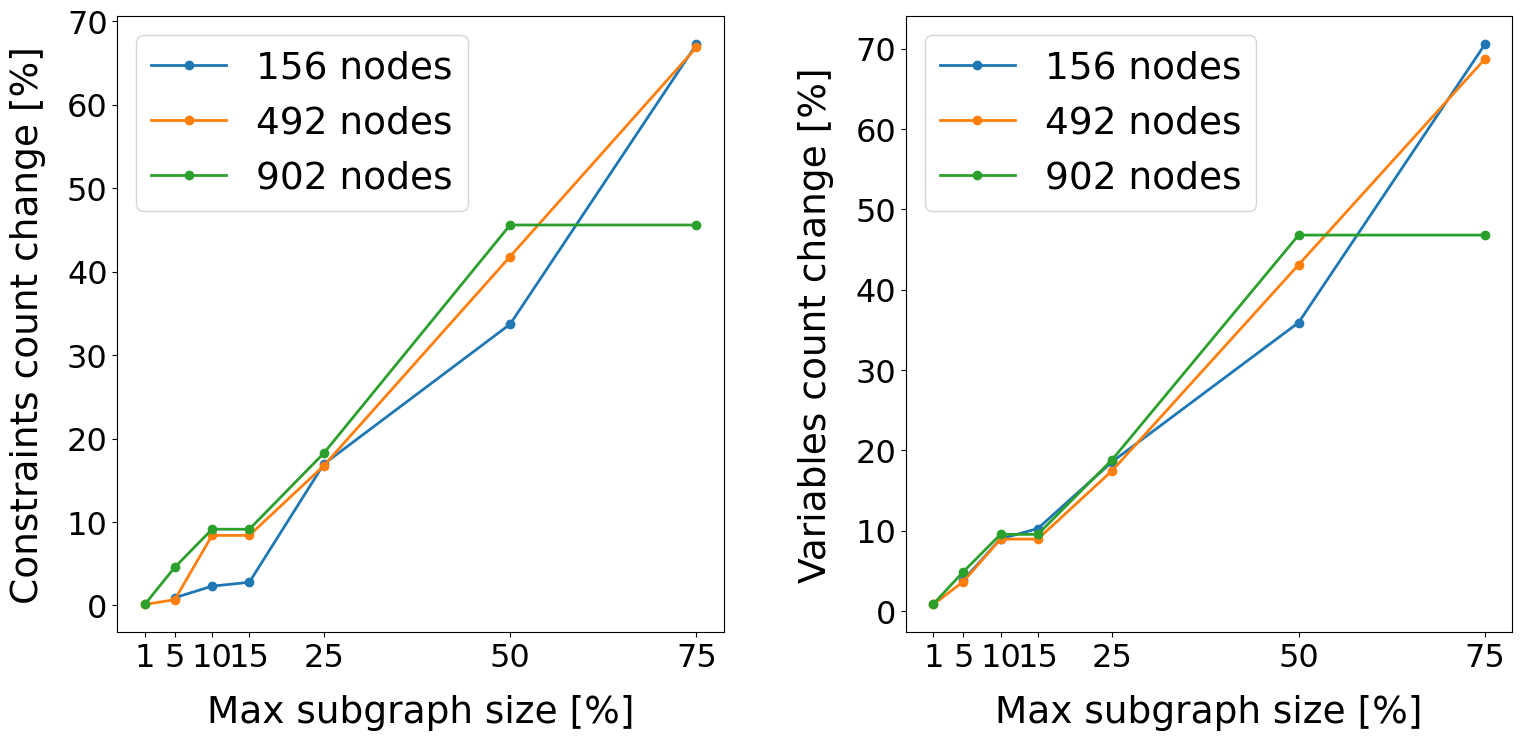}}\label{fig:mss_problem_size_genome}
	\quad
	\subfloat[Montage]{\includegraphics[width=\linewidth]{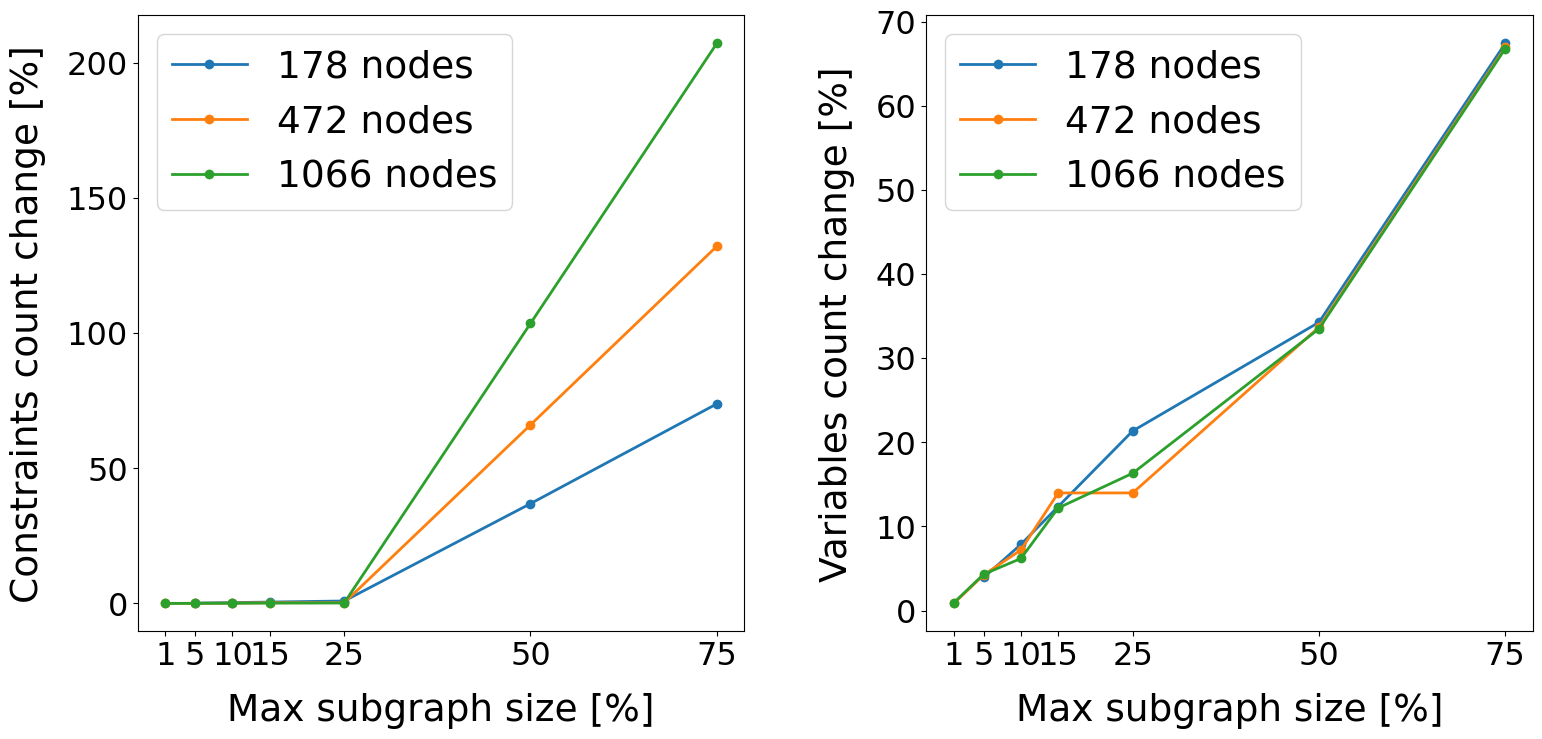}}
	\caption{Maximum subgraph size influence on the number of constraints and variables.}
	\label{fig:problem_size_mss} 
\end{figure}

Selected results are presented in Fig.~\ref{fig:problem_size_mss}. The SPWD algorithm applied to  1000Genome family instances reduced the problem size for each tested {\em max subgraph size} parameter value (Fig.~\ref{fig:problem_size_mss}a). Such behavior is reasonable because smaller graphs typically contain fewer paths and tasks. Similar results were observed for the Epigenomics, Seismology, SoyKB and SRA Search families (not shown in the figure).

Slightly different results were observed for the Montage and Cycles workflows. For Montage, as shown in Fig.\ref{fig:problem_size_mss}b, the number of variables decreased similarly to 1000Genome. However, the number of constraints increased significantly for the {\em max subgraph size} values above $50\%$ and this effect strengthened with the size of the workflow. The SPWD algorithm may still be profitable if the {\em max subgraph size} is small enough. Scheduling with values above $50\%$ will not provide any benefit. Similar trends were observed for the Cycles workflow family but were even more intensified (not shown in the figure).

\subsubsection{TTSP mapping influence}
\begin{figure}[htbp]
    \centering
    \subfloat[Montage]{\includegraphics[width=0.4\linewidth]{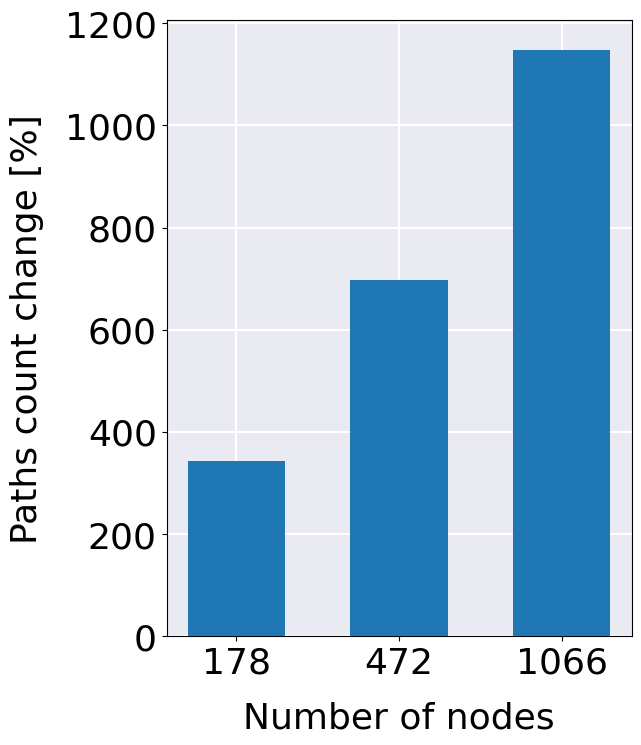}}
    \qquad
    \subfloat[Cycles]{\includegraphics[width=0.41\linewidth]{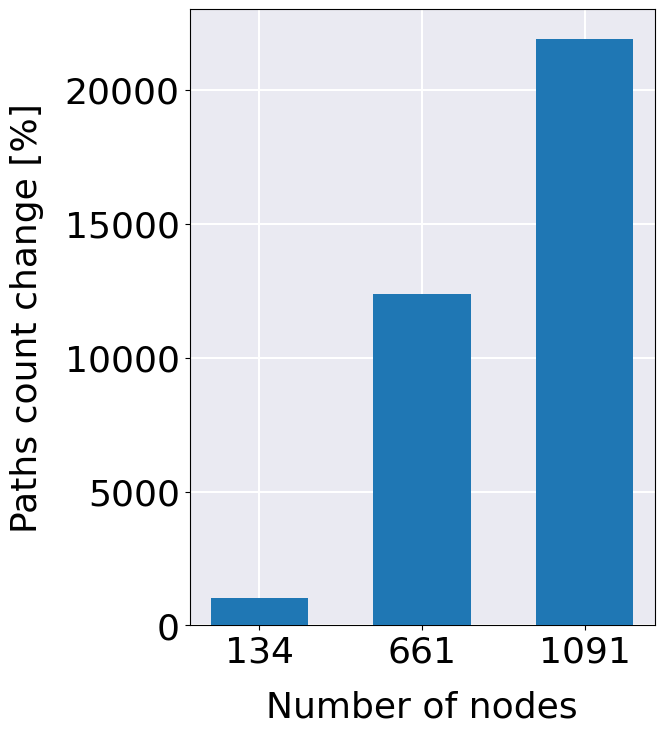}}
    \caption{Relative increase of the number of paths caused by mapping workflow to TTSP form.}
    \label{fig:ttsp_path_increase} 
\end{figure}

We further investigated the problem with the number of constraints for Montage and Cycles workflows (see Section~\ref{sec:mss_problem_size}). As presented in Fig.~\ref{fig:ttsp_path_increase}, for each workflow examined, the number of paths increased after the TTSP mapping, which explains the need for a small {\em max subgraph size} for these families to actually decrease the size of the problem.

\subsection{Usability for quantum solvers}
The most important part of our research was determining whether the decomposition algorithm improves the quality of scheduling large workflows with the CQM hybrid sampler.

\subsubsection{CQM capacity limits}\label{sec:mss_cqmcapasity}
To measure the capacity of the CQM solver, we directly solved WfCommons instances from different families, with various numbers of variables and constraints. The tests were performed in June $2024$. The collected results are shown in Tab.~\ref{tab:cqm_limits}. Each CQM result was compared with the Gurobi solver in terms of the cost increase (4th column of Tab.~\ref{tab:cqm_limits}). Huge degradation in CQM's performance occurred when the number of constraints exceeded $17,000$, where correct solutions could not be found.

Although we suspected that variables could also potentially cause solver degradation,  it did not seem to be the case since $32,715$ variables did not cause any significant scheduling cost increase.

\begin{table}[H]
    \centering
    \caption{Relative CQM solver solution quality  for selected WfCommons instances compared to the Gurobi result. Empty cells mean that CQM was unable to schedule the workflow and did not return the correct result.}\label{tab:cqm_limits}
    \begin{tabular}{|c|c|c|c|}
        \hline
        workflow family &variables count &constraints count &cost increase \\
        \hline
        Montage & 30 & 10 & 1 \\
        \hline
        Epigenomics & 6,985 & 1,743 & 1 \\
        \hline
        Cycles & 10,910 & 3,802 & 1 \\
        \hline
        1000Genome & 2,460 & 4,860 & 1.002 \\
        \hline
        Montage & 890 & 8,062 & 1.017 \\
        \hline
        1000Genome & 4,510 & 8,910 & 1.009 \\
        \hline
        Montage & 545 & 8,929 & 1.06 \\
        \hline
        Cycles & 32,715 & 11,403 & 1 \\
        \hline
        Montage & 525 & 17,001 & - \\
        \hline
        Montage & 1,550 & 25,846 & - \\
        \hline
        Montage & 2,360 & 46,744 & - \\
        \hline
    \end{tabular}
\end{table}

\subsubsection{Improving CQM capacity with SPWD}
Testing the SPWD algorithm with the CQM solver was performed using all Pegasus instances that were not solvable directly by CQM, i.e. four Montage workflow instances listed in Tab.~\ref{tab:cqm_problems}. 
For each instance, we manually selected the maximum subgraph size value, scheduled decomposed workflows with CQM, and compared the results with Gurobi, as presented in Tab.~\ref{tab:montage_instances}.
\begin{table}[H]
    \centering
    \caption{Problem sizes for scheduling on CQM.}\label{tab:cqm_problems}
    \begin{tabular}{|c|c|c|}
        \hline
        workflow &variables count &constraints count \\
        \hline
        Montage\_310 &1,550 &25,846 \\
        \hline
        Montage\_472 &2,360 &46,744 \\
        \hline
        Montage\_619 &3,095 &102,499 \\
        \hline
        Montage\_1066 &5,330 &181,366 \\
        \hline
    \end{tabular}
\end{table}
The SPWD algorithm allowed us to successfully schedule all the workflow instances that CQM could not solve previously. 
The third and fourth columns of Tab.~\ref{tab:montage_instances} show the schedule cost increase compared to the results obtained from the Gurobi solver, supported and unsupported by decomposition, respectively. 
The column marked as $s$ indicates the value of the parameter {\em max subgraph size}.
The scheduling cost obtained with SPWD was very similar for the CQM and Gurobi solvers (the difference did not exceed $1\%$ for Gurobi's benefit). The cost overhead caused by CQM supported by SPWD compared to the raw Gurobi result was also acceptable, and the highest increase observed was $15.8\%$ for Montage 619.
\begin{table}[H]
    \centering
    \caption{The cost increase for Montage workflows scheduled with the CQM using decomposition, compared with  Gurobi solver.}\label{tab:montage_instances}
    \begin{tabular}{|c|c|c|c|}
        \hline
        workflow & $s$ &$\frac{\text{CQM + SPWD}}{\text{Gurobi + SPWD}}$  [\%] &$\frac{\text{CQM + SPWD}}{\text{Gurobi (raw)}}$ [\%] \\
        \hline
        Montage\_310 &100 &0.0 &8.0 \\
        \hline
        Montage\_472 &150 &0.0 &1.4 \\
        \hline
        Montage\_619 &200 &0.7 &15.8 \\
        \hline
        Montage\_1066 &350 &0.1 &1.2 \\
        \hline
    \end{tabular}
\end{table}

\section{Summmary and Conclusions}
\label{sec:conclusions}
In this paper, we proposed a new SPWD workflow decomposition algorithm, based on the concept of TTSP graphs. The algorithm's behavior can be flexibly adjusted using the {\em max subgraph size} parameter, which allows it to be applied to a wide range of workflows. Moreover, the algorithm has been aligned with the QHyper framework, which allowed its seamless integration with various solvers. As a result, we were able to use the algorithm to overcome the current capacity limitations of the hybrid quantum-classical CQM solver and find schedules for real-life workflow instances with up to $181,366$ constraints.

The algorithm has been successfully applied to various workflows, and the highest cost increase was around $17.5\%$, when the scheduled workflow size decreased $100$ times. For most workflow families, any {\em max subgraph size} value effectively caused the problem size to decrease. However, the step of mapping to TTSP added many new paths to Montage and Cycles workflows, which required small values of {\em max subgraph size} to alleviate this effect.

Future work could involve introducing different parameters that control how workflows are divided, such as the maximum number of constraints. This would be more convenient to use, as its value could be established once, based on the capacity limits of a given solver, and be applied for every processed workflow. Another interesting research topic would be to apply SPWD to a pure quantum annealer. The capacity limitations of such solvers were signaled in \cite{krzhizhanovskaya_foundations_2020} and decomposition could improve scheduling results. However, this step would require an efficient and automatic method of setting penalty weights in the WSP QUBO function.
\section*{CRediT author statement}
K.R. conceptualized and supervised research; M.K. designed and implemented SPWD algorithm; J.Z. designed and implemented execution environment (QHyper) for experiments with CQM and Gurobi solvers as well as for integration with WfCommons format; M.K designed and performed the experiments; M.K. and K.R. analyzed the results; M.K. and J.Z. created the figures. All authors took part in writing the first draft of the manuscript as well as revised, edited, and approved the final version of the manuscript.
\section{Acknowledgments}
\noindent We would like to thank Tomasz Lamża, Mariusz Sterzel, Kacper Jurek for QHyper support and Mateusz Hurbol for CQM and WfCommons support. The research presented in this paper received partial funding from Polish Ministry of Science and Higher Education assigned to AGH University of Krakow. 
We gratefully acknowledge Polish high-performance computing infrastructure PLGrid (HPC Center: ACK Cyfronet AGH) for providing computer facilities and support within computational grant no. PLG\-/2024/\-017208.

\bibliographystyle{elsarticle-num}
\bibliography{references}

\end{document}